\documentclass[twocolumn,prl,npa,showpacs]{revtex4}
\usepackage{amsmath,bm}
\usepackage{graphicx}
\usepackage{ulem}
\usepackage{bm}
\usepackage{CJK}

\begin{document}
\begin{CJK*}{GBK}{}

\renewcommand{\thefootnote}{\fnsymbol{footnote}}
\renewcommand{\figurename}{Fig.}

\title{The second Born approximation of electron-argon elastic scattering in a Bichromatic laser field}

\author{Bin Zhou$^{1}$ , Ming-Yang Zheng$^{2}$ and Da-Yong Wen$^{3}$}
\affiliation{ $^1$Physics lab, nanxu college, Jiangsu University of
Science and Technology, Zhenjiang, Jiangsu 212003, People's Republic of China \\
 $^2$Department of Modern Physics, University of
Science and Technology of China, Hefei, Anhui 230026, People's
Republic of China\\$^3$Nanxu college, Jiangsu University of Science
and Technology, Zhenjiang, Jiangsu 212003, People's Republic of
China.}
\date{\today}

\begin{abstract}
We study the elastic scattering of atomic argon by electron in the
presence of a bichromatic laser field in the second Born
approximation. The target atom is approximated by a simple screening
potential and the continuum states of the impinging and emitting
electrons are described as Volkov states. We evaluate the S-matrix
elements numerically. The dependence of differential cross section
on the relative phase between the two laser components is presented.
The results obtained in the first and second Born approximation are
compared and analysed.
\end{abstract}

\pacs{34.80.Qb; 32.80.Wr; 34.50.Rk; 34.80.Bm}

\maketitle

The process of multiphoton free-free transitions (MFFT) was firstly
studied by Bunkin and Fedorov \cite{F.V. Bunkin}, and has attracted
much attention in physical community. Summaries of these
investigations were presented in the book by Mittleman \cite{M.H.
Mittleman} and some reviews \cite{C.J. Joachain,Fritz}. The
theoretical work treating the laser radiation classically with a
single frequency $\omega$, or some narrow band multi-mode
approximation has yielded prefect agreement with the experiments by
Weigngarshofer \cite{Weingartshofer}. With the development of laser
technology, the atomic and molecular processes assisted or induced
by powerful and new kinds of laser fields have been researched,
especially for the case by multicolor laser. Free-free transitions
in a powerful bichromatic laser field has become feasible
experimentally to coherently control the phase between the two
components of the radiation field. A considerable body of research
work has concentrated on the coherent phase control (CPC) of the
elastic stattering processes \cite{Sun07,D Nehari110,BA
deHarak10,Ghalim M,Zhang S T,Protopapas M,Kamiski,Varr,Zhang S T
and,Varr S and,Ehlotzky,Cionga A,Miloevic}. In such mentioned
papers, the multiphoton processes were treated in the first Born
approximation (FBA), while in this letter, we carried out the
calculation in the second-order Born approximation (SBA) and
compared the results with that in the FBA formation. Atomic units
$\hbar=m=e=1$ are used throughout.

Considering that in a laser beam the density of radiation quanta is
so large that the depletion of this beam by emitting or absorbing
quanta from it is negligible, then the laser field is treated
classically in our model. Hence, the bichromatic laser field is
described as a classical electromagnetic field with the fundamental
frequency $\omega$ and its second harmonic 2$\omega$, i.e.,
$\boldsymbol{\mathcal {E}}(t)=\boldsymbol{\mathcal
{E}}_\circ[\sin\omega t+\sin(2\omega t+\varphi)]$, where
$\boldsymbol{\mathcal {E}}_\circ$ is the electric field amplitude
and the relative phase $\varphi$ can be arbitrarily changed.

The target atom is described by a screening potential
\cite{Salvant}:
\begin{equation}
V(r)=-\frac{\textrm{Z}}{r}\sum_{i=1}^3A_i\exp(-\alpha_ir),
\label{eq1}
\end{equation}
where $r$ denotes the position of the electron with respect to the
nucleus, and $\textrm{Z}$ is the nuclear charge number. For argon,
$A_1=2.1912$, $A_2=-2.8252$, $A_3=1-A_1-A_2$, $\alpha_1=5.5470$,
$\alpha_2=4.5687$, and $\alpha_3=2.0446$.

The scattering matrix for the laser-assisted free-free transition in
the second Born approximation reads:
\begin{equation}
S^{<2>}_{fi}=S^{(1)}_{fi}+S^{(2)}_{fi}, \label{eq2}
\end{equation}
where
\begin{equation}
S^{(1)}_{fi}=-i\langle\chi_{\boldsymbol{k}_f}|V|\chi_{\boldsymbol{k}_i}\rangle.
\label{eq3}
\end{equation}
Here $\chi_{\boldsymbol{k}_i}$ and $\chi_{\boldsymbol{k}_f}$ are the
states of the electrons in the initial and final channels, described
by the Volkov wave function:
\begin{eqnarray}
\chi_{\boldsymbol{k}_{i,f}}&=&\exp(i\boldsymbol{k}_{i,f}\cdot r)\exp
\left[-iE_{i,f}t-\frac{i}{\omega^2}\boldsymbol{k}_{i,f}\cdot\boldsymbol{\mathcal
{E}}_\circ\sin\omega t\right]\nonumber\\&&
\exp\left[-\frac{i}{4\omega^2}\boldsymbol{k}_{i,f}\cdot\boldsymbol{\mathcal
{E}}_\circ\sin(2\omega t+\varphi)\right], \label{eq4}
\end{eqnarray}
where ${\boldsymbol{k}}_{i,f}$ are the wave vectors of incident and
scattered electrons, and $E_{i,f}$ are the corresponding kinetic
energies.

\begin{figure}[htbp]
\centering
\includegraphics[width=16cm,bb= 0 0 842 595]{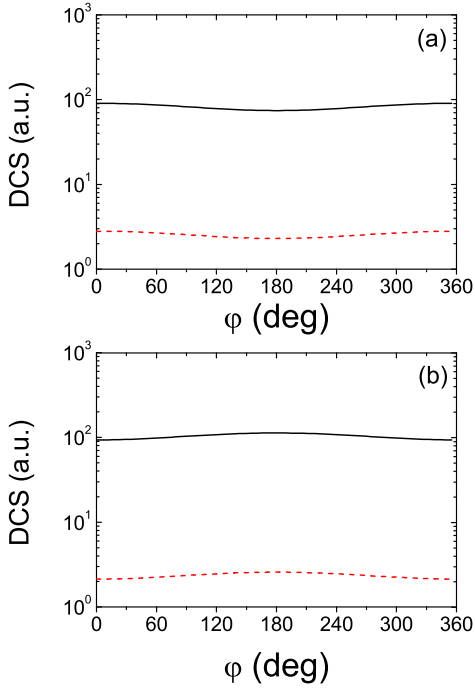}
\caption{\label{fig:epsart}(a) The DCS versus $\varphi$ for elastic
electron-argon scattering with emission of one photon ($l=1$). The
scattered angle of emitting electron is $\theta=13^\circ$. The
kinetic energies of the incident electron is $E_i=9.5 \
e\textrm{V}$. Solid curve: SBA results; dashed curve: FBA results.
(b) The same as (a), but for the case with one photon absorption
($l=-1$).}
\end{figure}

\begin{figure}[htbp]
\centering
\includegraphics[width=16cm,bb= 0 0 842 595]{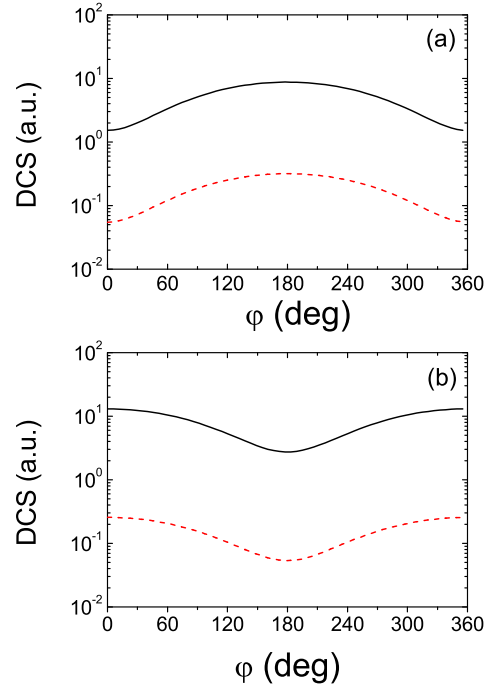}
\caption{\label{fig:epsart}(a) The parameters are the same as Fig. 1
except for the number of the exchanged photons. For (a), $l=2$,
while for (b), $l=-2$. Solid curve: SBA results; dashed curve: FBA
results.}
\end{figure}

\begin{figure}[htbp]
\centering
\includegraphics[width=16cm,bb= 0 0 842 595]{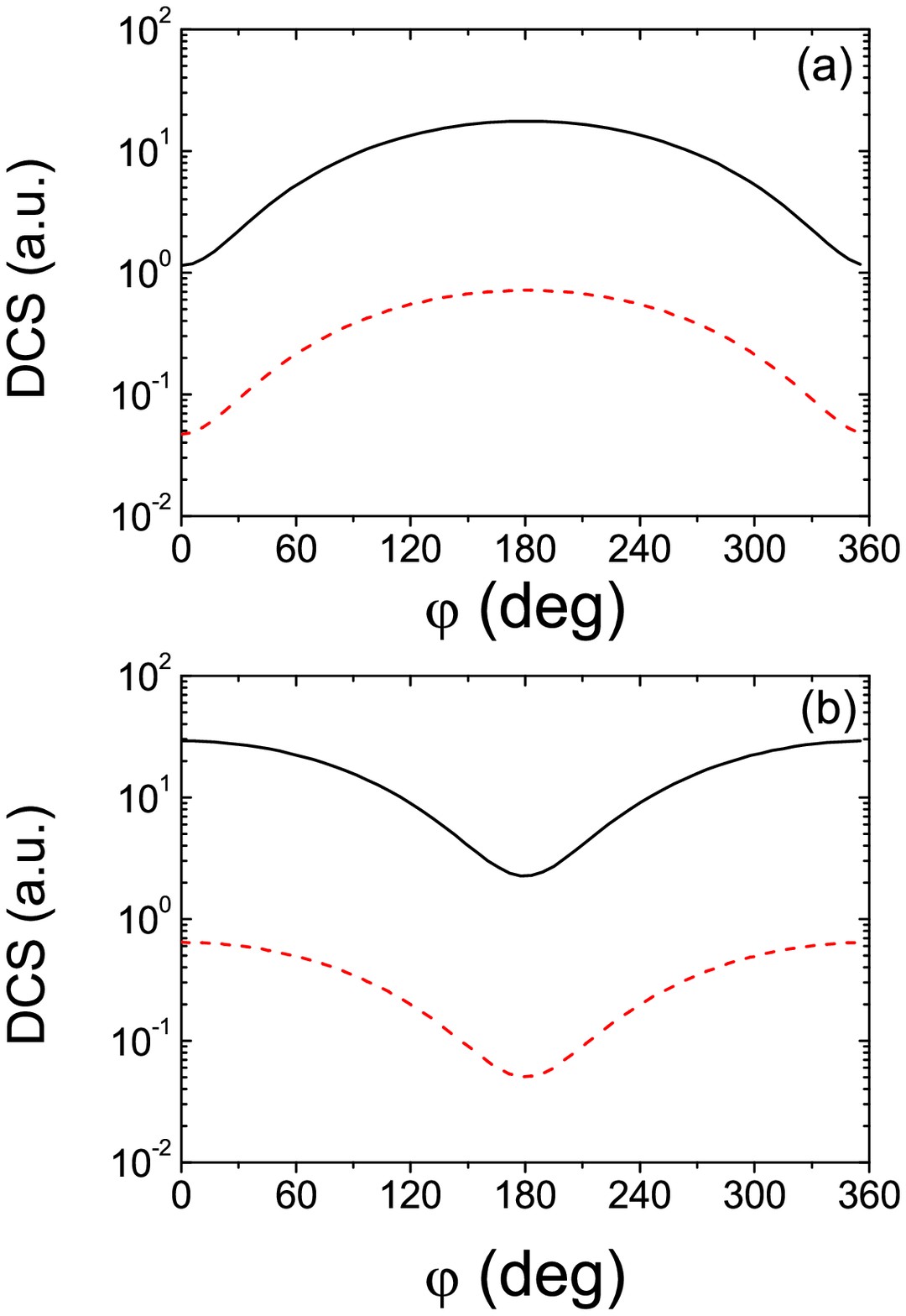}
\caption{\label{fig:epsart} The parameters are the same as Fig. 2
except for the impact energy. In this figure, the energy of the
incoming electron is $E_i=19.5 \ e\textrm{V}$. Solid curve: SBA
results; dashed curve: FBA results.}
\end{figure}

Using the potential of Eq. (\ref{eq1}) and the wave functions in Eq.
(\ref{eq4}), the first term of the $S_{fi}^{<2>}$ can be recast as:
\begin{equation}
S^{(1)}_{fi}=-2\pi i\sum_{l} T_{fi}^{(1)}(l)\delta(E_f-E_i+l\omega).
\label{eq5}
\end{equation}
$T_{fi}^{(1)}(l)$ is the ionization amplitude accompanying the
exchange of $l$ photons with the laser field ($l > 0$ for emission,
and $l < 0$ for absorption),
\begin{equation}
T_{fi}^{(1)}(l)=B_l(\lambda,\frac{1}{4}\lambda,\varphi)V(\boldsymbol{k}_{f,i}),
\label{eq6}
\end{equation}
in which
\begin{equation}
V(\boldsymbol{k}_{f,i})=\int d{\bm r}e^{-{\mathrm{i}}({\bm k}_f-{\bm
k}_i)\cdot {\bm r}}V(r). \label{eq7}
\end{equation}
Clearly it is just the Fourier transformation of the potential. The
other term has the form:
\begin{equation}
B_l(\lambda,\frac{1}{4}\lambda,\varphi)=\sum_{n=-\infty}^\infty
J_{l-2n}(\lambda)J_n(\frac{1}{4}\lambda)\exp(-{\mathrm{i}}n\varphi),
\label{eq8}
\end{equation}
which is the generalized Bessel function with $\lambda=({\bm
k}_f-{\bm k}_i)\cdot \boldsymbol{\mathcal {E}}_0/\omega^2$. $J_n$ is
the ordinary Bessel function.

The second term of SBA matrix is:
\begin{eqnarray}
S_{fi}^{(2)}&=&-i\int d\boldsymbol{r}\int
d\boldsymbol{r'}\int_{-\infty}^\infty dt\int_{-\infty}^\infty
dt'\chi_{\boldsymbol{k}_f}^\ast(\boldsymbol{r},t)\nonumber\\&&V(r)
G(\boldsymbol{r},t;\boldsymbol{r}',t')
V(r')\chi_{\boldsymbol{k}_i}(\boldsymbol{r'},t)\nonumber,
\label{eq9}
\end{eqnarray}
where $G(\boldsymbol{r},t;\boldsymbol{r}',t')$ is the Green
function:
\begin{equation}
G(\boldsymbol{r},t;\boldsymbol{r}',t')=-\frac{i}{(2\pi)^3}\int
d\boldsymbol{k}\chi_{\boldsymbol{k}}(\boldsymbol{r},t)\chi_{\boldsymbol{k}}^\ast(\boldsymbol{r},t)u(t-t').
\label{eq10}
\end{equation}
Here $u(t-t')$ is the step function.

Using Eq. (\ref{eq4}) and Eq. (\ref{eq10}), we obtain:
\begin{eqnarray}
S_{fi}^{(2)}=-2\pi i \sum_{l}
T_{fi}^{(2)}(l)\delta(E_f-E_i+l\omega), \label{eq11}
\end{eqnarray}
where the photon-number-resolved transition amplitude can be
expressed as:
\begin{eqnarray}
T_{fi}^{(2)}(l)&=&\frac{1}{(2\pi)^3}\sum_{m}\int
d\boldsymbol{k}\frac{1}{E_i-E-m\omega+i\eta}\nonumber\\&&
V(\boldsymbol{k}_f,\boldsymbol{k})
V(\boldsymbol{k},\boldsymbol{k}_i)
B_m(\lambda_1,\frac{1}{4}\lambda_1,\varphi)\nonumber\\&&
B_{l-m}(\lambda_2,\frac{1}{4}\lambda_2,\varphi), \label{eq12}
\end{eqnarray}
where $\eta$ is a small positive quantity.

So we can obtain the SBA scattering amplitude:
\begin{equation}
T_{fi}^{<2>}(l)=T_{fi}^{(1)}(l)+T_{fi}^{(2)}(l). \label{eq13}
\end{equation}

The differential cross sections (DCS) for the net exchange of $l$
photons between the colliding system and the bichromatic laser field
can be described as:
\begin{eqnarray}
\frac{d\sigma}{d\Omega}=(\frac{1}{2\pi})^2\frac{k_f}{k_i}|T_{fi}^{<2>}(l)|^2.
\label{eq14}
\end{eqnarray}

For numerical calculation, we studied the dependence of differential
cross section for electron-argon atom scattering on the relative
phase angle $\varphi$ between the two laser components under the
geometry of the experiment by Weingartshofer \cite{Weingartshofer}.
The angle between the polarization vector $\boldsymbol{\mathcal
{E}}$ and the momentum of the incident electron ${\bf k}_i$ is
$\phi=38^\circ$, the momentum ${\bf k}_f$ of the scattered electron
is in the plane defined by the polarization vector
$\boldsymbol{\mathcal {E}}$ and ${\bf k}_i$. The bichromatic laser
parameters are $\omega=0.117 \ e\textrm{V}$ and $ \mathcal
{E}_0=2.7\times10^8\ \textrm{V} /cm$.

In Fig. 1, we show the DCS versus the phase angle $\varphi$. The CPC
effects are not very apparent for one photon exchange process. The
distribution of the controlling effect is centered on
$\varphi=180^\circ$. Such phenomenon is more distinct for the
processes with more than one photons transfer which could be seen
clearly in the following two figures. This may attribute to the fact
that the generalized Bessel function satisfies the relationship
\cite{Fritz}: $B_{-l}(a,b,\varphi)=(-1)^lB_l^*(a,b,\varphi-\pi)$.
Moreover, the results have some prominent improvements in the second
Born approximation. This indicates that the intermediat states have
significant contribution to the collision processes.

Figure 2 displays the DCS versus $\varphi$ for two photons transfer.
As is depicted in the figure, the DCS for $l=\pm 2$ are much smaller
than $l=\pm 1$. This demonstrates that the probability of two
photons transfer between laser field and the colliding system is
small. With the increase of $|l|$ (not presented here), the DCS
becomes smaller and smaller. For the case with two photons emission
($l=2$), more electrons are emitted at $\theta=13^\circ$ when the
phase difference between the laser components is near
$\varphi=180^\circ$ where the CPC effect reaches the maximum; while
for $l=-2$, the situation is inverse.  The second-order corrections
are still distinct, and for different $\varphi$, the contribution of
the intermediate states are almost the same.

In Fig. 3, we show the influence of the impact energy on the
collision process. It is evident that the DCS becomes larger when
the energy of the incident electron increases and the CPC effects
are more prominent.

In summary, the electron-argon elastic scattering in a bichromatic
laser field is investigated in the second Born approximation. The
dependence of DCS on the relative phase is analysed. The DCS are
significantly improved in the second Born approximation. We
attribute it to the contributions of the intermediate states.

This work is supported by the National Nature Science Foundation of
China under Grants No. 10874169 and No. 10674125, and the National
Basic Research Program of China under Grants No. 2007CB925200 and
No. 2010CB923301.

\end{CJK*}
\end{document}